\documentclass[11pt]{article}

\usepackage{amsmath,amssymb}
\usepackage{graphicx}
\usepackage{graphics}
\usepackage{dcolumn}
\usepackage{bm}
\usepackage{afterpage}
\usepackage[mathscr]{euscript}
\usepackage{color}

\usepackage[a4paper,left=1.75cm,right=1.75cm,top=3.0cm,bottom=3.0cm]{geometry}

\numberwithin{equation}{section}

\begin{document}

\allowdisplaybreaks

\title{Another Mass Gap in the BTZ Geometry?}

\vskip1cm
\author{Sean Stotyn and Robert B. Mann\\ \\ \it{Department of Physics and Astronomy, University of Waterloo,}\\ \it{Waterloo, Ontario, Canada, N2L 3G1}\\ \\
                   \small{smastoty@sciborg.uwaterloo.ca, rbmann@sciborg.uwaterloo.ca}}

\date{}
\maketitle

\begin{abstract}

We attempt the construction of perturbative rotating hairy black holes and boson stars, invariant under a single helical Killing field, in $2+1$-dimensions to complete the perturbative analysis in arbitrary odd dimension recently put forth in \cite{Stotyn:2011ns}.  Unlike the higher dimensional cases, we find evidence for the non-existence of hairy black holes in $2+1$-dimensions in the perturbative regime, which is interpreted as another mass gap, within which the black holes cannot have hair.  The boson star solutions face a similar impediment in the background of a conical singularity with a sufficiently high angular deficit, most notably in the zero-mass BTZ background where boson stars cannot exist at all.  We construct such boson stars in the AdS$_3$ background as well as in the background of conical singularities of periodicities $\pi,\frac{2\pi}{3},\frac{\pi}{2}$.

\end{abstract}

\section{Introduction}

One of the important contributions that Stuart Dowker made to theoretical physics was in his study of boson gases in curved space-time \cite{Dowker:1978md}.  In obtaining the  high temperature expansion \cite{Dowker:1988jw}  and chemical potential \cite{Dowker:1989gp}  for the free energy of a massive, non-conformally coupled,
ideal scalar gas in a static spacetime, he laid the foundations for understanding not only the quantum properties of black hole radiation
\cite{Mann:1996ze,Mann:1996bi}
 but other novel phenomena as well, such as Bose-Einstein condensation as a symmetry-breaking effect in curved spacetime
 \cite{Toms:1993ic}.   Following in this tradition, we shall demonstrate here that there are still interesting new things to learn by studying scalar fields in curved space-time. In particular, we shall provide evidence for a new mass gap for black holes in (2+1) dimensions, within which the black holes cannot have hair.

Recently, there has been interest sparked in the subject of asymptotically anti-de Sitter soliton configurations and hairy black hole solutions which are invariant under a single Killing vector field, first constructed in \cite{Dias:2011at} for $D=5$, then extended in \cite{Stotyn:2011ns} to $D=7,9,11$, and further extended in $D=5$ to matter fields with Maxwell charges \cite{Dias:2011tj}.  The soliton configurations in these analyses are known as boson stars and represent matter configurations of (un)charged massless scalar fields whose self-gravitation is balanced by their asymptotic charges, i.e. centrifugal repulsion from rotation and/or charge repulsion from Maxwell fields.  The hairy black holes, on the other hand, represent the stable end state of a superradiant instability of spinning black holes; the scalar field undergoes superradiant scattering off the horizon, mining rotational energy from the black hole, then is reflected by the AdS boundary back toward the horizon.  This process continues until the scalar field has extracted all the rotational energy it can and the resulting configuration is a lump of scalar field co-rotating with the black hole.  These solutions are invariant under a single helical Killing vector field and circumvent the stationarity theorems \cite{Hawking:1971vc,Hollands:2006rj,Moncrief:2008mr}, which state that a stationary solution must have at least 2 Killing vectors, because they are not stationary but are instead harmonic in time.  The construction of these hairy black hole solutions crucially depends on a judicious choice of cohomogeneity-1 metric and scalar field ansatz such that the matter stress tensor shares the same symmetries as the metric.  The resulting equations of motion then form a set of coupled second order ordinary differential equations instead of a system of coupled partial differential equations.   

This construction is currently only known to work in odd spacetime dimensions and has been carried out in all odd dimensions of interest in string theory except in $D=3$.  In much the same way that black holes in 2+1 dimensions have very different properties than their higher dimensional analogues, the analysis of \cite{Stotyn:2011ns}, valid for arbitrary odd dimension $D\ge5$, was inappropriate for addressing the $D=3$ case.  This distinction is essentially an artifact of the zero-mass black hole background in 2+1 being distinct from the AdS$_3$ vacuum.  The aim of this paper is to complete this perturbative analysis in odd dimensions by considering a massless scalar field minimally coupled to Einstein gravity with negative cosmological constant in 2+1 dimensions.  Hairy black holes in this theory correspond to scalar hair co-rotating with a BTZ black hole, although arguably the most interesting result we find in $D=3$ is the absence of hairy black holes in the perturbative regime, $r_+\ll\ell$.   This can be understood via an important feature of the effective potential specific to $D=3$: gravitational attraction and centrifugal repulsion enter the effective potential at the same power of $r$.  This means that in a certain range of black hole parameters gravitational attraction dominates and the effective potential is strictly attractive, while in another range rotational repulsion dominates and the effective potential is strictly repulsive; only the latter case can support scalar hair.  For small black holes, there does not exist a stable configuration of hair around them because the centrifugal repulsion from rotation cannot overcome the gravitational attraction.  This suggests that there is another mass gap between the zero-mass geometry and a minimum size black hole, above which the rotational repulsion of the scalar field will be strong enough to balance its gravitational attraction to the black hole leading to configurations with stable scalar hair.  Finding this threshold requires solving the full set of equations of motion numerically and is not treated in this paper.

The remainder of the paper is outlined as follows: in section 2 we introduce the ansatz for the metric and the scalar field and give the resulting equations of motion.  In section 3 we construct boson stars as perturbations around global AdS$_3$ as well as perturbations around orbifolded AdS$_3$.  We explicitly show by a coordinate rescaling that orbifolded AdS$_3$ spacetimes are identical to a subset of asymptotically AdS conical singularities, i.e. ``orbifolded AdS$_3$" actually corresponds to states in the mass gap between pure AdS$_3$ and the zero-mass BTZ black hole.  We thus focus on constructing solutions in the background of generic conical singularities. However we find the equations cannot be solved in closed form for arbitrary deficit angle, so we give explicit results for a few specific choices.  We continue in section 4 by showing that there are no perturbative hairy black holes in 2+1 dimensions and finally we conclude in section 5 with a discussion of the issues raised throughout the paper.

\section{Ansatz and Equations of Motion}

We begin with 3 dimensional Einstein gravity with negative cosmological constant, $\Lambda=-\frac{1}{\ell^2}$, minimally coupled to a complex scalar field
\begin{equation}
S=\frac{1}{16\pi}\int{d^3x\sqrt{-g}\left(R+\frac{2}{\ell^2}-2\big|\nabla\Pi\big|^2\right)}.\label{eq:action}
\end{equation}
To draw a connection to rotating BTZ black holes without scalar hair, we consider the metric and scalar field ansatz
\begin{equation}
ds^2=-f(r)g(r)dt^2+\frac{dr^2}{f(r)}+r^2\big(d\phi-\Omega(r) dt\big)^2\label{eq:metric}
\end{equation}
\begin{equation}
\Pi=\Pi(r) e^{-i\omega t+in\phi} \label{eq:ScalarField}
\end{equation}
so that the rotating BTZ solution of \cite{Banados:1992wn} is given by
\begin{equation}
f(r)=\frac{r^2}{\ell^2}-M+\frac{J^2}{4r^2}, \quad\quad\quad g(r)=1, \quad\quad\quad \Omega(r)=\frac{J}{2r^2}, \quad\quad\quad \Pi(r)=0 \label{eq:BTZmetric}
\end{equation}
where $M$ and $J$ are the ADM mass and angular momentum respectively.  The angular coordinate has range $\phi\in[0,2\pi]$ and the radial coordinate has range $r\in[r_+,\infty)$, where $r_+$ is the outermost zero of $f(r)$.  We note that the scalar field (\ref{eq:ScalarField}) is single valued if we choose an orbifolded space such that $\phi\in\left[0,\frac{2\pi}{n}\right]$ where $n\in{\mathbb Z}^+$.  However, we will show in section 3 that orbifolded AdS$_3$ spaces are, in fact, non-orbifolded AdS$_3$ spaces in disguise.

The stress tensor for the scalar field takes the form
\begin{equation}
T_{ab}=\left(\partial_a\Pi^*\partial_b\Pi+\partial_a\Pi\partial_b\Pi^*\right)-g_{ab}\left(\partial_c\Pi\partial^c\Pi^*\right),\label{eq:Tab}
\end{equation}
which has the same symmetries as the metric (\ref{eq:metric}), although the scalar field does not: the metric (\ref{eq:metric}) is invariant under $\partial_t$, $\partial_\phi$ while the scalar field (\ref{eq:ScalarField}) is only invariant under the combination
\begin{equation}
K=\partial_t+\frac{\omega}{n}\partial_\phi. \label{eq:KV}
\end{equation}
Therefore, any solution with nontrivial scalar field will only be invariant under the single helical Killing vector field given by (\ref{eq:KV}).  For reasons that will become apparent in the following sections, we choose $n=1$ above.  

The equations of motion resulting from the action (\ref{eq:action}) are $G_{ab}-\frac{1}{\ell^2}g_{ab}=T_{ab}$ and $\nabla^2\Pi=0$.  Plugging the ansatz (\ref{eq:metric}) and (\ref{eq:ScalarField}) into the equations of motion yields a system of coupled second order ODEs
\begin{equation}
f''+f'\left(\frac{3}{r}-\frac{g'}{2g}\right)+\frac{fg'}{g}\left(\frac{1}{r}-\frac{g'}{2g}\right)+\frac{8\Pi'\Pi}{r}+\frac{4\Pi^2}{fg}(\omega-\Omega)^2+\frac{4\Pi^2}{r^2}-\frac{8\Pi^2\Omega'r}{fg}(\omega-\Omega)-\frac{8}{\ell^2}=0
\end{equation}
\begin{equation}
g''+g'\left(\frac{2f'}{f}+\frac{1}{r}\right)-\frac{8\Pi^2}{f^2}(\omega-\Omega)^2+\frac{8r\Pi^2\Omega'}{f^2}(\omega-\Omega)-\frac{8g\Pi'\Pi}{rf}=0\label{eq:gEq}
\end{equation}
\begin{equation}
\Omega''+\frac{4\Pi^2}{fr^2}(\omega-\Omega)+\Omega'\left(\frac{3}{r}-\frac{g'}{2g}\right)=0\label{eq:OmegaEq}
\end{equation}
\begin{equation}
\Pi''+\frac{\Pi'(f^2gr^2)'}{2f^2gr^2}+\frac{\Pi}{f^2g}(\omega-\Omega)^2-\frac{\Pi}{fr^2}=0\label{eq:PiEq}
\end{equation}
where $\Xi=\Pi^2-\frac{r^2}{L^2}$ and a $'$ denotes differentiation with respect to $r$.  In addition to these second order ODEs, the Einstein equations further impose two first order ODEs in the form of constraint equations, $C_1=0$ and $C_2=0$. Explicitly, these are
\begin{equation}
C_1=\frac{r}{f}(f^2g)'+4g\Xi+r^4\Omega'^2
\end{equation}
\begin{equation}
C_2=\frac{\Pi^2(\omega-\Omega)^2}{f^2g}+\Pi'^2+\frac{r^2\Omega'^2}{4fg}+\frac{f'}{2fr}+\frac{\Pi^2}{fr^2}-\frac{1}{f\ell^2}.
\end{equation}
We note that the above equations of motion cannot be obtained simply by taking $h=1$ and $n=1$ in Ref. \cite{Stotyn:2011ns} so the $D=3$ case indeed requires separate consideration.

 Finally, we consider null geodesics with 3-velocity $k^\alpha=(\dot t,\dot r,\dot \phi)$ in the BTZ background, where a dot represents a derivative with respect to some affine parameter $\lambda$.  If we define the Killing vectors $\xi=\partial_t$ and $\zeta=\partial_\phi$ then the conserved quantities along the geodesics are
\begin{equation}
E=\left(f(r)-r^2\Omega^2(r)\right)\dot t+r^2\Omega(r)\dot\phi, \quad\quad\quad\quad L=-r^2\Omega(r)\dot t+r^2\dot\phi.
\end{equation}
This leads to an energy equation of the form $0=\dot r^2+V_{eff}$ where the effective potential is
\begin{equation}
V_{eff}=\frac{L^2}{\ell^2}-E^2-\frac{L}{r^2}\big(ML-JE\big).  \label{eq:EffectivePotential}
\end{equation}
There are a few comments to be made about this effective potential.  First, the attractive term proportional to $M$ enters at the same power of $r$ as the repulsive term proportional to $J$.  Thus the potential is either attractive if $ML>JE$, meaning that the null rays invariably get sucked into the black hole, or repulsive if $JE>ML$, meaning that the null rays can be reflected off the centrifugal barrier.  We will discuss this more in section 4.  Secondly, the effective potential is constant if $M=J=0$, which corresponds to the zero-mass BTZ background; we will show in section 3 that a boson star cannot exist in this background.  Lastly, if $M<0$ and $J=0$, corresponding to naked conical singularities as well as AdS$_3$, the effective potential is always repulsive.  Only when the effective potential is repulsive can we expect to find a stable nontrivial scalar field supported; indeed we will show that boson stars exist in the backgrounds of conical singularities. However under the model presented in this section, perturbative hairy black holes are not possible because the effective potential is attractive and cannot support scalar hair. 

\section{Perturbative Boson Stars}\label{Boson}
In this section, we construct boson star solutions as perturbations around global AdS$_3$ as well as around conical singularities.  We begin by explicitly showing that orbifolded AdS$_3$ spaces do not exist in the sense that they are non-orbifolded AdS$_3$ spaces in disguise.  Because of this, we extend our analysis to constructing boson star solutions as perturbations around conical singularities of arbitrary deficit angle.  This treatment is special to $D=3$ because such conical singularities correspond to states in the mass gap between pure AdS$_3$ and the zero-mass BTZ background.  The expansion is carried out in orders of the scalar field condensate parameter $\epsilon$ and we give results up to order $\epsilon^6$.  As a perturbative construction, these results will only be valid for small energies and angular momenta.

\subsection{No Orbifolded AdS$_3$ Spaces}

Orbifolded spaces are obtained by taking the quotient of an angular orbit by a discrete subgroup, the simplest example of which is ${\mathbb R}^k/\mathbb{Z}_n$.  This generically creates a conical fixed point at the center of the orbifold and it explicitly breaks supersymmetry by requiring fermions to have anti-periodic boundary conditions asymptotically.  The orbifold further induces localized closed string tachyon condensation around the fixed point, which leads to the decay of the spacetime via a ``dilaton pulse," as was shown in Ref. \cite{Adams:2001sv} for the orbifold ${\mathbb C}^2/{\mathbb Z}_n$, with $n$ necessarily odd to avoid closed string tachyons in the bulk; in that case, the orbifolded plane decays to the flat plane.  Following the analysis of \cite{Adams:2001sv}, the authors of Ref. \cite{Hikida:2007pr} concluded that orbifolded AdS$_3$ decays to the AdS$_3$ vacuum via a dilaton pulse, however they restricted their attention to orbifolds AdS$_3/{\mathbb Z}_n$ with $n$ odd.  This restriction is unnecessary as we will now show that AdS$_3/{\mathbb Z}_n$ is identical to a subset of conical singularities in the mass gap between pure AdS$_3$ and the zero-mass BTZ geometry.  Therefore, by the analysis of \cite{Hikida:2007pr}, any conical singularity ought to undergo decay to AdS$_3$ since they all share the same asymptotic structure.

Consider the orbifolded AdS$_3$ geometry written in standard form
\begin{equation}
ds^2=-\left(\frac{R^2}{\ell^2}+1\right)dT^2+\frac{dR^2}{\left(\frac{R^2}{\ell^2}+1\right)}+R^2d\varphi^2
\end{equation}
where $\varphi\in\left[0,\frac{2\pi}{n}\right]$ such that $n\in{\mathbb Z}^+$.  Now perform the following coordinate transformation:
\begin{equation}
T=\frac{t}{n}, \quad\quad\quad R=n r,\quad\quad\quad \varphi=\frac{\phi}{n} \label{eq:CoordTrans}
\end{equation}
such that $\phi\in[0,2\pi]$.  The metric now takes the form
\begin{equation}
ds^2=-\left(\frac{r^2}{\ell^2}+\frac{1}{n^2}\right)dt^2+\frac{dr^2}{\left(\frac{r^2}{\ell^2}+\frac{1}{n^2}\right)}+r^2d\phi^2
\end{equation}
which is a conical singularity with $M=-\frac{1}{n^2}$ within the mass gap $-1< M < 0$.  In fact, it was noted in \cite{Hikida:2007pr} that AdS$_3/{\mathbb Z}_n$ has the same asymptotic structure for all $n$ and now we see precisely why this is the case: AdS$_3/{\mathbb Z}_n$ are simply conical singularities written in poorly scaled coordinates.  This subtlety is special to AdS$_3$ and does not occur, for instance, in the ${\mathbb C}^2/{\mathbb Z}_n$ case considered in \cite{Adams:2001sv} since there is no way to turn the cone into a plane by scaling the radial coordinate.  In the rest of this paper, then, we will focus on the more generic case of constructing boson stars as perturbations around conical singularities of arbitrary deficit angle.

Before continuing, however, we take the opportunity to extend the previous discussion to  orbifolded BTZ black holes: consider such a black hole, given by:
\begin{equation}
ds^2=-f(R)dT^2+\frac{dR^2}{f(R)}+R^2\big(d\varphi-\Omega(R)dT\big)^2
\end{equation}
where $\varphi\in\left[0,\frac{2\pi}{n}\right]$ and the metric functions are given by $f(R)=\frac{R^2}{\ell^2}-M+\frac{J^2}{4R^2}$ and $\Omega(R)=\frac{J}{2R^2}$.  Now perform the coordinate transformation (\ref{eq:CoordTrans}), as well as rescale the mass and angular momenta by
\begin{equation}
J=n^2 j, \quad\quad\quad M=n^2 m
\end{equation}
so that the metric takes the form
\begin{equation}
ds^2=-\tilde f(r)dt^2+\frac{dr^2}{\tilde f(r)}+r^2\big(d\phi-\tilde\Omega(r)dt\big)^2
\end{equation}
where $\phi\in[0,2\pi]$ and the metric functions are given by $\tilde f(r)=\frac{r^2}{\ell^2}-m+\frac{j^2}{4r^2}$ and $\tilde\Omega(r)=\frac{j}{2r^2}$.  Comparing with (\ref{eq:BTZmetric}), this is simply the standard BTZ metric with mass $m$ and angular momentum $j$.

\subsection{Boson Star Origin}

Boson stars are smooth, horizonless geometries, which means that all metric functions must be regular at the origin: the angular deficit of the conical singularity -- or lack thereof --  must be preserved when the scalar field is added, which restricts the lowest order terms of the metric functions.  To find the boundary condition on $\Pi$, we multiply (\ref{eq:PiEq}) by $r^2$ and note that $\Pi$ must vanish at the origin in order to yield consistent equations of motion.   Thus, the boundary conditions at the boson star origin take the form
\begin{eqnarray}
\left.f\right|_{r \rightarrow 0} = \frac{1}{\mu^2} + \mathcal{O}(r^2), \quad 
\left.g\right|_{r \rightarrow 0} = g(0)+\mathcal{O}(r), \quad 
\left.\Omega\right|_{r \rightarrow 0} = \Omega(0)+\mathcal{O}(r), \quad 
\left.\Pi\right|_{r \rightarrow 0} = \mathcal{O}(r^\alpha), \label{eq:OriginBC}
\end{eqnarray}
where we have introduced $M=-\frac{1}{\mu^2}$ for convenience and $\alpha>0$.  Here $M$ is the ADM mass of the background appearing in (\ref{eq:BTZmetric}) such that AdS$_3$ corresponds to $\mu=1$, and conical singularities correspond to $\mu>1$ with the ``orbifolds" of the previous subsection being the subset $\mu\in{\mathbb Z}^+$.  The zero-mass BTZ background is given by the limit $\mu\rightarrow\infty$.

\subsection{Asymptotic Boundary Conditions}

In order to simplify the asymptotic boundary conditions, we first make note of a residual gauge freedom.  It is straightforward to show that the transformation
\begin{eqnarray} \label{PsiGaugeFreedom}
\phi \rightarrow \phi + \lambda t, \quad\quad \Omega \rightarrow \Omega + \lambda, \quad\quad \omega \rightarrow \omega - \lambda
\end{eqnarray}
for some arbitrary constant $\lambda$, leaves both the metric (\ref{eq:metric}) and scalar field (\ref{eq:ScalarField}) unchanged.  We use this gauge invariance to pick a frame which is not rotating at infinity, i.e. we use it to set $\Omega \rightarrow 0$ in the limit $r \rightarrow \infty$.  In this limit the boundary conditions will be the same for the boson star and the black hole, in particular they will asymptote to the BTZ metric (\ref{eq:BTZmetric}) with higher order multipole corrections:
\begin{align}
\left.f\right|_{r \rightarrow \infty}= \frac{r^2}{\ell^2} -{\mathscr M}  + \mathcal{O}(r^{-2}), \quad\>\> 
\left.g\right|_{r \rightarrow \infty} &= 1  + \mathcal{O}(r^{-4}), \quad\>\>
\left.\Omega\right|_{r \rightarrow \infty} = \frac{\mathscr J}{2r^2} + \mathcal{O}(r^{-4}), \label{eq:AsymBC}\\ 
\left.\Pi\right|_{r \rightarrow \infty}&= \frac{\epsilon \ell^2}{r^2} + \mathcal{O}(r^{-3}).\nonumber
\end{align}
where $\mathscr M$ and $\mathscr J$ are the ADM mass and angular momentum of the full solution respectively.  The boundary condition on $\Pi$ is set by requiring it to be normalizable.  Here and in what follows, $\epsilon\ll1$ provides a dimensionless measure of the amplitude of the scalar field, 

\subsection{Constructing Perturbative Solutions}\label{PertBS}

We start by expanding our fields in terms of the scalar field amplitude as follows:
\begin{equation}
F(r,\epsilon)=\sum_{i=0}^m{F_{2i}(r)\epsilon^{2i}}\quad\quad\quad \Pi(r,\epsilon)=\sum_{i=0}^m{\Pi_{2i+1}(r)\epsilon^{2i+1}}\quad\quad\quad \omega(\epsilon)=\sum_{i=0}^m{\omega_{2i}\epsilon^{2i}} \label{eq:FieldExpansion}
\end{equation}
where $F=\{f,g,h,\Omega\}$ is shorthand for each of the metric functions in (\ref{eq:metric}).  The metric functions are expanded in even powers of $\epsilon$ while the scalar field is expanded in odd powers.  This allows a perturbative expansion as follows: start at $m=0$ with the desired background and introduce a nontrivial scalar field by solving (\ref{eq:PiEq}) to order $\epsilon$.  At next order, $\Pi_1(r)$ sources the gravitational fields $F_2(r)$, which in turn affect the scalar field via $\Pi_3(r)$.  The perturbative solution can, in principle, be obtained by this bootstrapping procedure up to arbitrary order, $m$. However we also must expand the frequency in even powers of $\epsilon$  because at  linear order the frequency is determined by the scalar field alone, whereas at higher orders  the back reacted gravitational fields induce nontrivial frame-dragging effects, which in turn affect the rotation of the scalar field.  In practice, these corrections to $\omega$ are found by imposing the boundary conditions.

We choose a conical singularity of arbitrary deficit angle as our background:
\begin{equation}
f_0=\frac{r^2}{\ell^2}+\frac{1}{\mu^2},\quad\quad g_0=1,\quad\quad \Omega_0=0.
\end{equation}
In this background, the most general massless scalar field solution to (\ref{eq:PiEq}) which is consistent with the asymptotic boundary conditions (\ref{eq:AsymBC}) is given by
\begin{equation}
\Pi_1(r)=\frac{r^\mu\ell^{2} }{(r^2+\ell^2/\mu^2)^{1+\frac{\mu}{2}}}{_2F_1}\left[\frac{\mu}{2}\left(1+\frac{2}{\mu}-\omega \ell\right),\frac{\mu}{2}\left(1+\frac{2}{\mu}+\omega \ell\right);2;\frac{\ell^2/\mu^2}{r^2+\ell^2/\mu^2}\right]\label{eq:Pi1}
\end{equation}
where $_2F_1$ is the hypergeometric function.  Now in order to satisfy the boundary conditions at the origin (\ref{eq:OriginBC}) we must further restrict $\omega$ to
\begin{equation}
\omega\ell=1+\frac{2}{\mu}+2k,\quad\quad\quad\quad k=0,1,2,...
\end{equation}
where the non-negative integer, $k$, describes the various possible radial modes of the scalar field.  We choose the mode $k=0$ as this corresponds to the ground state, in which case (\ref{eq:Pi1}) simplifies to
\begin{equation}
\Pi_1(r)=\frac{r^\mu\ell^2 }{(r^2+\ell^2/\mu^2)^{1+\frac{\mu}{2}}}. \label{eq:Pi1}
\end{equation}

Proceeding up the perturbative ladder, we insert (\ref{eq:Pi1}) and the expansion (\ref{eq:FieldExpansion}) into the equations of motion, and solve for order $\epsilon^2$.  In general, the solutions contain two constants of integration, which are then uniquely fixed by the boundary conditions.  These fields, $F_2(r)$, are then inserted into the equation of motion for $\Pi(r)$ to find the $\epsilon^3$ correction to the scalar fields.  At the order $\epsilon^2$ level, the equations of motion can still be solved easily for arbitrary values of $\mu$, however this does not remain true at order $\epsilon^3$ and higher.  Thus, we give explicit results for the few select values $\mu=1,2,3,4$; these values yield concise expressions for the fields, whereas other non-integer values of $\mu$ are either difficult to solve or yield expressions that are too cumbersome to warrant writing down explicitly.  Calculating the field expansions up to order $\epsilon^6$, we find 
\begin{equation}
f_2(r)=-\frac{2\mu^2r^{2\mu}\big((2+\mu)r^2+(1+\mu)\ell^2/\mu\big)}{(1+\mu)\big(r^2+\ell^2/\mu^2\big)^{1+\mu}},\label{eq:f2}
\end{equation}
\begin{equation}
g_2(r)=\frac{2\mu^4}{1+\mu}\left(-2+\frac{r^{2\mu}\big(2r^4+2(2+\mu)r^2\ell^2/\mu^2+(1+\mu)\ell^4/\mu^3\big)}{\big(r^2+\ell^2/\mu^2\big)^{2+\mu}}\right),\label{eq:g2}
\end{equation}
\begin{equation}
\Omega_2(r)=\frac{\mu^4\big(2\big(r^2+\ell^2/\mu^2\big)^{1+\mu}-r^{2\mu}\big(2r^2+(2+\mu)\ell^2/\mu^2\big)\big)}{\ell(1+\mu)\big(r^2+\ell^2/\mu^2\big)^{1+\mu}},\label{eq:Omega2}
\end{equation}
\begin{equation}
\Pi_3(r)=\frac{\ell^4r^\mu\Pi_{\mu;3}}{\big(r^2+\ell^2/\mu^2\big)^{\frac{3(2+\mu)}{2}}},\label{eq:Pi3}
\end{equation}
\begin{equation}
f_4(r)=-\frac{r^{2\mu}f_{\mu;4}}{(r^2+\ell^2/\mu^2)^{3+2\mu}},
\end{equation}\label{eq:f4}
\begin{equation}
g_4(r)=-\frac{\ell^{6}g_{\mu;4}}{(r^2+\ell^2/\mu^2)^{4+2\mu}},
\end{equation}\label{eq:g4}
\begin{equation}
\Omega_4(r)=\frac{\ell\Omega_{\mu;4}}{(r^2+\ell^2/\mu^2)^{3+2\mu}},
\end{equation}\label{eq:Omega4}
\begin{equation}
\Pi_5(r)=\frac{\ell^{4}r^\mu\Pi_{\mu;5}}{(r^2+\ell^2/\mu^2)^{\frac{5(2+\mu)}{2}}},\label{eq:Pi5}
\end{equation}
where the fields $\{f_{\mu;n},g_{\mu;n},\Omega_{\mu;n},\Pi_{\mu;n}\}$ are simple polynomials in $r$ for the case of $\mu\in{\mathbb Z}^+$ but are more complicated functions for arbitrary choices of $\mu$.  These fields are catalogued in Appendix B for $\mu=1,2,3,4$.

We note from (\ref{eq:f2}) that there is a condition on the existence of a boson star in a conical singularity background with a large deficit angle.  This is because sufficiently far from $r=0$, in the regime of $\mu\gg1$, $f(r)$ takes the form
\begin{equation}
f(r)\approx \frac{r^2}{\ell^2}+\frac{1}{\mu^2}-2\epsilon^2\mu^2.
\end{equation}
For a fixed value of $\mu\gg1$, there is a range of sufficiently large values of $\epsilon\ll1$ such that $f$ turns negative at a finite radius, signaling the formation of a horizon.  Hence, for a conical singularity of sufficiently high deficit angle, there is a maximum amplitude to the scalar field, beyond which the boson star undergoes gravitational collapse.   However in the limit $\mu\rightarrow\infty$, corresponding to the zero-mass BTZ background, \emph{any} amount of scalar field causes a horizon to form.  Recall that the zero-mass background has a constant effective potential, meaning that the scalar field is able to make its way to $r=0$.  Now consider a radially ingoing null ray: $0=-\frac{r^2}{\ell^2}\dot t^2+\frac{\dot r^2}{r^2/\ell^2}$.  Rearranging and integrating from a finite radius to $r=0$ we note that it takes an infinite coordinate time for the null ray to reach $r=0$, meanwhile the proper volume of spacetime is shrinking exponentially toward $r=0$.  This allows a piling up of the scalar field in a vanishing volume around the origin, eventually forming a horizon as a result.  This suggests that the zero-mass BTZ background is perturbatively unstable toward forming black holes.   Even though in the next section we show that there are no perturbative hairy black holes in $D=3$, stable hairy black holes cannot be found by constructing boson stars that form horizons because boundary conditions must be imposed at the horizon and these are not guaranteed to be satisfied \emph{a priori}.

Finally, since these boson stars are invariant under the Killing vector field (\ref{eq:KV}), they must satisfy the first law of thermodynamics, which follows from a Hamiltonian derivation of the first law \cite{Wald:1993ki}.  Since boson stars are horizonless objects, they have zero entropy and their thermodynamics are determined completely by their energy and angular momentum.  The first law in this case takes the form
\begin{equation}
\mathrm{d}{\mathscr M}=\omega \mathrm{d}{\mathscr J}.
\end{equation}
These thermodynamic quantities are easily found using the asymptotic boundary conditions (\ref{eq:AsymBC}), in conjunction with the field expansions (\ref{eq:Pi3})-(\ref{eq:Pi5}), and are catalogued in Appendix A for $\mu=1,2,3,4$.

\section{No Perturbative Hairy Black Holes}

In this section we briefly discuss the problem of constructing rotating hairy BTZ black holes in this perturbative construction.  There are now two expansion parameters: the scalar field amplitude, $\epsilon\ll1$, and the dimensionless measure of the size of the black hole $r_+/\ell\ll1$.  To search for small hairy black holes, we perform a double expansion of our fields as follows
\begin{align}
F(r,\epsilon,r_+/\ell)=\sum_{i=0}^m\sum_{j=0}^n{F_{2i,2j}(r)\epsilon^{2i}\left(\frac{r_+}{\ell}\right)^{2j}}, \quad&\quad\quad \Pi(r,\epsilon,r_+/\ell)=\sum_{i=0}^m\sum_{j=0}^n{\Pi_{2i+1,2j}(r)\epsilon^{2i+1}\left(\frac{r_+}{\ell}\right)^{2j}},\\
\omega(\epsilon,r_+/\ell)={}&\sum_{i=0}^m\sum_{j=0}^n{\omega_{2i,2j}\epsilon^{2i}\left(\frac{r_+}{\ell}\right)^{2j}}, \label{eq:DoubleFieldExpansion}
\end{align}
where we start with a background consisting of a small black hole, which has the gravitational fields $\sum_{j=0}^{n}F_{0,2j}(r)(r_+/\ell)^{2j}$.  Actually, we must expand the fields and solve for them in the outer region, $r\gg\ell$, as well as in the near-horizon region $r_+\le r\ll\infty$ and match the two solutions in the intermediate region $r_+\ll r\ll\infty$, a process called matched asymptotic expansion.  For full details of this procedure, we refer the reader to ref. \cite{Stotyn:2011ns}.  For the purposes of this discussion, however, we simply note that the two asymptotic solutions for a given field have four constants of integration, two of which are uniquely fixed by the boundary conditions and the other two uniquely fixed by the matching procedure.

The equation of motion (\ref{eq:OmegaEq}) imposes the physical condition coming from superradiance that the scalar field and black hole must be co-rotating, $\omega=\Omega_H$, where $\Omega_H$ is the angular velocity of the horizon.  In terms of $r_+$ and $\omega$, the BTZ metric functions take the form
\begin{equation}
f(r)=\frac{r^2}{\ell^2}-\frac{r_+^2}{\ell^2}\big(1+\omega^2\ell^2\big)+\frac{r_+^4\omega^2}{r^2}, \quad\quad g(r)=1, \quad\quad \Omega(r)=\frac{r_+^2\omega}{r^2},
\end{equation}
which can be expanded in powers of $r_+/\ell$ to the desired order by virtue of the order $\epsilon^0$ expansion for $\omega$.  We also need the boundary conditions at the horizon, which take the form
\begin{align}
&\left.f\right|_{r_+} =\mathcal{O}(r-r_+),   &\left.g\right|_{r_+} = g(r_+)+\mathcal{O}(r-r_+),\\
&\left.\Omega\right|_{r_+} = \omega+\mathcal{O}(r-r_+),  & \left.\Pi\right|_{r_+} = \Pi(r_+)+\mathcal{O}(r-r_+), \label{eq:HorizonBC}
\end{align}
where the condition $\Omega(r_+)=\omega$ is imposed by the equations of motion.  Physically this is the statement that the black hole and scalar field must be co-rotating.

From the considerations in the last section, there are further restrictions on the fields we must take into account.  We already noted that a boson star cannot be supported in the zero-mass BTZ background.  This means that in the $r_+\rightarrow 0$ limit we must recover the zero-mass background without any scalar field, which implies $F_{2n,0}(r)=0$ and $\Pi_{2n-1,0}(r)=0$ for $n=1,2,3,...$.  We also saw in the previous section that the zero mode of the frequency in the zero-mass BTZ background ($\mu\rightarrow\infty$) is $\omega_{0,0}=1/\ell$; indeed if we ignore this and keep $\omega_{0,0}$ arbitrary we find that the equations of motion impose algebraic equations forcing all the scalar field coefficient functions to vanish except if and only if $\omega_{0,0}=1/\ell$.  

This exhausts the available information we must exploit to look for solutions describing small BTZ black holes with co-rotating scalar hair.  Plugging in the appropriate field expansions (\ref{eq:DoubleFieldExpansion}) to the equations of motion, keeping only the non-zero coefficient functions, we immediately run into a problem when attempting to add a nontrivial scalar field to a small black hole background.  
Solving the equations for $\Pi_{1,2}$ in the far and near-horizon regions, the matching procedure then forces $\Pi_{1,2}$ to vanish identically.  Explicitly, the near and far region solutions for $\Pi_{1,2}$ that satisfy the respective asymptotic and near-horizon boundary conditions are
\begin{align}
&\Pi^{out}_{1,2}=C_1\frac{\ell^2}{r^2},  &\Pi^{in}_{1,2}=\frac{C_2}{\sqrt{z^2-\ell^2}}K_1\left[\sqrt{\frac{2\ell^3\omega_{0,2}}{z^2-\ell^2}}\right],
\end{align}
where $K_1[x]$ is the modified Bessel function of the second kind and $z=\ell r/r_+$ is the near-horizon radial coordinate chosen so that $z\ge\ell\gg r_+$.  The above solution for $\Pi^{in}_{1,2}$ vanishes at the horizon, while the other homogeneous solution to the second order ODE is proportional to $I_1[x]$, which diverges at the horizon and hence doesn't satisfy the boundary conditions; $I_1$ is the modified Bessel function of the first kind with $x=\sqrt{2\ell^3\omega_{0,2}/(z^2-\ell^2)}$.  We have taken the sign of $\omega_{0,2}$ to be negative, \emph{i.e.} $\omega=1/\ell-\omega_{0,2}r_+^2/\ell^2+...$, because making the replacement $\omega_{0,2}\rightarrow-\omega_{0,2}$ turns $\big\{I_1[x],K_1[x]\big\}\rightarrow\big\{J_1[x],Y_1[x]\big\}$ (Bessel functions of the first and second kind respectively), both of which yield solutions that diverge and oscillate infinitely often as the horizon is approached.  The boundary conditions at the horizon then force $\omega_{0,2}$ to come with a minus sign and the scalar field satisfying such boundary conditions is given above.  Taking the small-$r$ limit of $\Pi^{out}_{1,2}$ is trivial, while the large-$z$ limit of $\Pi^{in}_{1,2}$ yields
\begin{equation}
\Pi^{in}_{1,2}\rightarrow \frac{C_2}{\sqrt{2\ell^3\omega_{0,2}}}+{\cal O}\left(\frac{r_+^2}{\ell^2}\right).
\end{equation}
The matching procedure now requires that
\begin{equation}
C_1\frac{\ell^2}{r^2}=\frac{C_2}{\sqrt{2\ell^3\omega_{0,2}}},
\end{equation}
which is only possible if $C_1=C_2=0$, meaning that $\Pi_{1,2}$ vanishes identically.
Similar results hold for $\Pi_{1,2n}$ for $n=1,2,3,...$, meaning that our perturbative approach is failing.

Small ($r_+\ll\ell$) BTZ black holes in this construction have an attractive effective potential, hence the scalar field cannot rotate fast enough to balance the gravitational attraction.   That we cannot find hairy black holes in the perturbative regime is indicative of this fact.  As a consequence, there must be a minimum size of BTZ black hole such that a co-rotating scalar field will have \emph{just enough} rotation to balance the gravitational attraction.  This represents another mass gap in the BTZ geometry, separating a ``no-hair" phase from a ``hairy" phase.  Unfortunately, our perturbative procedure is ill-equipped to find where this phase transition takes place and we must resort to numerical methods.  We leave this work for future consideration.    

Finally, we note that we find similar non-existence results if we consider higher order perturbative modes, \emph{i.e.} $n>1$ for the scalar field ansatz (\ref{eq:ScalarField}).  This can be expected by considering the effective potential (\ref{eq:EffectivePotential}): if a stable hairy black hole were to exist, such a configuration would have a lower angular momentum $J$ relative to its mass $M$, making the effective potential \emph{more} attractive.  We also note that allowing self-interaction for the scalar field, or including Maxwell charges may be able to provide enough repulsion to make configurations with stable hair possible: it should be understood that the mass gap we are inferring corresponds to BTZ black holes with non-interacting massless scalar hair.

\section{Conclusion}

We have considered a massless scalar field minimally coupled to Einstein gravity with a negative cosmological constant to construct boson star solutions as perturbations around AdS$_3$ as well as perturbations around conical singularities of arbitrary deficit angle.   We have also shown that AdS$_3/{\mathbb Z}_n$ spaces don't exist in the usual sense insofar as they are actually non-orbifolded AdS$_3$ spaces written in poorly scaled coordinates.  This shows that empty AdS$_3/{\mathbb Z}_n$ is the same as an asymptotically AdS conical singularity of mass $M=-1/n^2$, which motivated our more generic consideration of boson stars around conical singularities of arbitrary deficit angle.  The gravitational and scalar fields were expanded in powers of the dimensionless asymptotic amplitude of the scalar field, $\epsilon$, in such a way as to provide a bootstrapping procedure for consistently building the solutions.  These boson star solutions are smooth and horizonless geometries representing soliton configurations of the scalar field where self-gravitation is exactly balanced by the centrifugal force of rotation.  These solitonic configurations are invariant under a single helical Killing vector field and have been verified to satisfy the first law of thermodynamics as objects of zero entropy.  For backgrounds consisting of conical singularities of sufficiently high deficit angle, i.e. ``steep" cones, there is a range of $\epsilon$ sufficiently large enough that the boson star will collapse to form a black hole: this is because the concentration of the scalar field in the cone is enough to cause a horizon to form.  The zero-mass BTZ background is the limit of a conical singularity with a periodicity of zero -- or a deficit angle of $2\pi$ -- and in this background, any amount of scalar field will cause a horizon to form.

We noted previously that conical singularities are unstable to decay to the AdS$_3$ vacuum via a dilaton pulse caused by closed string tachyon condensation around the tip.  Since this is a local phenomenon and the presence of the scalar field has vanishing effect at the conical tip, we expect our boson star solutions to be equally unstable to decay.  However, we have imposed reflecting boundary conditions on our solutions so we expect by time-symmetry that the dilaton pulse will be reflected off the boundary and re-form the conical singularity.  Thus, in the absence of a boson star, an AdS$_3$ conical singularity spacetime with reflecting boundary conditions ought to exhibit a ``breathing" dilaton pulse.  It is unclear how the presence of the boson star will affect this process: at the linear level the scalar field is non-interacting but at higher perturbative levels the situation is more complicated due to frame dragging and back-reaction.  One possibility is that the boson star provides a damping mechanism for the dilaton pulse, mediating a true decay down to AdS$_3$.  Such questions are open for future considerations.

The most provocative result of this paper is the non-existence of perturbative hairy black holes in 2+1 dimensions.  From superradiance considerations, the scalar field must be co-rotating with the black hole horizon; it is precisely this rotation that balances the gravitational attraction toward the black hole, yielding a stable end state configuration.  In the case of perturbative BTZ black holes, the effective potential for null particles is an attractive sink, hence the gravitational attraction of the scalar field toward the black hole is greater than the centrifugal repulsion from rotation.  This means there will be a minimum size of BTZ black hole such that balance between gravitational attraction and centrifugal repulsion will be possible and scalar hair will be supported.  This implies that there is a mass gap separating BTZ black holes without hair from those with hair.  It would be interesting to have an understanding of this phenomenon from an AdS/CFT perspective.  We leave this analysis, and the study of hairy BTZ black holes via numerical methods, for future work.

\vskip .5cm
\centerline{\bf Acknowledgements}
\vskip .2cm

This work was made possible by the Natural Sciences and Engineering Research Council of Canada

\appendix

\section{Thermodynamic Quantities}

\subsubsection*{$\mu=1$:}

\begin{equation}
{\mathscr M}_1=-1+3\epsilon^2+\frac{531}{100}\epsilon^4+{\cal O}(\epsilon^6), \quad\quad\quad {\mathscr J}_1=\ell\big(\epsilon^2+\frac{1853}{900}\epsilon^4+{\cal O}(\epsilon^6)\big),\nonumber
\end{equation}
\begin{equation}
\omega_1\ell=3-\frac{26}{15}\epsilon^2-\frac{89863}{23625}\epsilon^4+{\cal O}(\epsilon^6)\nonumber
\end{equation}

\subsubsection*{$\mu=2$:}

\begin{equation}
{\mathscr M}_2=-\frac14+\frac{32}{3}\epsilon^2+\frac{6869504}{33075}\epsilon^4+{\cal O}(\epsilon^6), \quad\quad\quad {\mathscr J}_2=\ell\big(\frac{16}{3}\epsilon^2+\frac{3864832}{33075}\epsilon^4+{\cal O}(\epsilon^6)\big),\nonumber
\end{equation}
\begin{equation}
\omega_2\ell=2-\frac{1024}{105}\epsilon^2-\frac{2855223296}{12733875}\epsilon^4+{\cal O}(\epsilon^6)\nonumber
\end{equation}

\subsubsection*{$\mu=3$:}

\begin{equation}
{\mathscr M}_3=-\frac19+\frac{45}{2}\epsilon^2+\frac{2654469}{1568}\epsilon^4+{\cal O}(\epsilon^6), \quad\quad\quad {\mathscr J}_3=\ell\big(\frac{27}{2}\epsilon^2+\frac{8766279}{7840}\epsilon^4+{\cal O}(\epsilon^6)\big),\nonumber
\end{equation}
\begin{equation}
\omega_3\ell=\frac{5}{3}-\frac{177}{7}\epsilon^2-\frac{8540255967}{3923920}\epsilon^4+{\cal O}(\epsilon^6)\nonumber
\end{equation}

\subsubsection*{$\mu=4$:}

\begin{equation}
{\mathscr M}_4=-\frac1{16}+\frac{192}{5}\epsilon^2+\frac{21129863168}{2858625}\epsilon^4+{\cal O}(\epsilon^6), \quad\quad\quad {\mathscr J}_4=\ell\big(\frac{128}{5}\epsilon^2+\frac{45817348096}{8575875}\epsilon^4+{\cal O}(\epsilon^6)\big),\nonumber
\end{equation}
\begin{equation}
\omega_4\ell=\frac{3}{2}-\frac{24064}{495}\epsilon^2-\frac{280779333197824}{26804509875}\epsilon^4+{\cal O}(\epsilon^6)\nonumber
\end{equation}

\section{Boson Star Fields}

\subsubsection*{$\mu=1$:}

\begin{align}
\Pi_{1;3}={}&\frac{1}{20}\big(\ell^2+r^2\big)\big(25\ell^2+56r^2\big)\nonumber\\
f_{1;4}={}&\frac{1}{100}\big(500 \ell^8 + 2350 \ell^6 r^2 + 4065 \ell^4 r^4 + 2630 \ell^2 r^6 + 531 r^8 \big)\nonumber\\
g_{1;4}={}&\frac{1}{150}\big(479 \ell^6 + 2724 \ell^4 r^2 + 5475 \ell^2 r^4 + 3440 r^6 \big)\nonumber\\
\Omega_{1;4}={}&\frac{1}{1800}\big(3742 \ell^8 + 14420 \ell^6 r^2 + 18450 \ell^4 r^4 + 9265 \ell^2 r^6 + 1853 r^8 \big)\nonumber\\
\Pi_{1;5}={}&\frac{1}{42000}\big(167265 \ell^{10} + 1050234 \ell^8 r^2 + 2510476 \ell^6 r^4 + 2768514 \ell^4 r^6 + 1352781 \ell^2 r^8 + 215554 r^{10}\big)\nonumber
\end{align}

\subsubsection*{$\mu=2$:}

\begin{align}
\Pi_{2;3}={}&\frac{1}{2520}\big(127 \ell^6 + 1984 \ell^4 r^2 + 14016 \ell^2 r^4 + 25728 r^6\big)\nonumber\\
f_{2;4}={}&\frac{1}{33075}\big(13335 \ell^{10} + 246400 \ell^8 r^2 + 1941100 \ell^6 r^4 + 7253904 \ell^4 r^6 + 11786432 \ell^2 r^8 + 6869504 r^{10}  \big)\nonumber\\
g_{2;4}={}&\frac{1}{529200}\big(2911 \ell^{10} + 93152\ell^8 r^2 + 1267168 \ell^6 r^4 + 9671424 \ell^4 r^6 + 35199360 \ell^2 r^8 + 43868160 r^{10} \big)\nonumber\\
\Omega_{2;4}={}&\frac{1}{1058400}\big(15283 \ell^{12} + 427924 \ell^{10} r^2 + 4791248 \ell^8 r^4 + 28152320 \ell^6 r^6 + 81000192 \ell^4 r^8\nonumber\\
& + 108215296 \ell^2 r^{10} + 61837312 r^{12} \big)\nonumber\\
\Pi_{2;5}={}&\frac{1}{39118464000}\big(258208199 \ell^{14} + 8150510752 \ell^{12} r^2 + 118176310304 \ell^{10} r^4 + 965394161920 \ell^8 r^6 \nonumber\\
&+ 4747429502720 \ell^6 r^8 + 12916379693056 \ell^4 r^{10} + 16957261774848 \ell^2 r^{12} + 7982908440576 r^{14}\big)\nonumber
\end{align}

\subsubsection*{$\mu=3$:}

\begin{align}
\Pi_{3;3}={}&\frac{1}{2204496}\big(1781 \ell^8 + 79065 \ell^6 r^2 + 1397898 \ell^4 r^4 + 15567066 \ell^2 r^6 + 
 48026520 r^8\big)\nonumber\\
f_{3;4}={}&\frac{1}{10287648}\big(99736 \ell^{12} + 4821810 \ell^{10} r^2 + 95874786 \ell^8 r^4 + 1092674772 \ell^6 r^6 + 6375205602 \ell^4 r^8\nonumber\\
& + 16947240147 \ell^2 r^{10} +17415971109 r^{12} \big)\nonumber\\
g_{3;4}={}&\frac{1}{168743146320}\big(249197 \ell^{14} + 22427730 \ell^{12} r^2 + 908323065 \ell^{10} r^4 + 21552666300 \ell^8 r^6\nonumber\\
&+ 330952717620 \ell^6 r^8 + 3448113919488 \ell^4 r^{10} + 17273628770580 \ell^2 r^{12} + 29730935304000 r^{14} \big)\nonumber\\
\Omega_{3;4}={}&\frac{1}{8332994880}\big(72946 \ell^{16} + 5908626 \ell^{14} r^2 + 212710536 \ell^{12} r^4 + 4378112046 \ell^{10} r^6 + 56042421750 \ell^8 r^8\nonumber\\
& + 458756168724 \ell^6 r^{10} + 2068179179004 \ell^4 r^{12} + 4658760078039 \ell^2 r^{14} + 4658760078039 r^{16} \big)\nonumber\\
\Pi_{3;5}={}&\frac{1}{364849681247251200}\big(628976914474 \ell^{18} + 56180440407060 \ell^{16} r^2 + 2255721444469890 \ell^{14} r^4\nonumber\\
& + 54923291791686360 \ell^{12} r^6 + 891183411545758695 \ell^{10} r^8 + 9801625653251321526 \ell^8 r^{10} \nonumber\\
&+ 73508902705618324245 \ell^6 r^{12} + 326433568800486019500 \ell^4 r^{14} + 726108769399804640550 \ell^2 r^{16}\nonumber\\
& + 610535114808723574200 r^{18}\big)\nonumber
\end{align}

\subsubsection*{$\mu=4$:}

\begin{align}
\Pi_{4;3}={}&\frac{1}{973209600}\big(6859 \ell^{10} + 652704 \ell^8 r^2 + 25824000 \ell^6 r^4 + 543047680 \ell^4 r^6 + 8330280960 \ell^2 r^8\nonumber\\
& + 36427530240 r^{10}\big)\nonumber\\
f_{4;4}={}&\frac{1}{23417856000}\big(2640715 \ell^{14} + 264520872 \ell^{12} r^2 + 11204911872 \ell^{10} r^4 + 259233863680 \ell^8 r^6\nonumber\\
& + 3893619671040 \ell^6 r^8 + 
 31044622090240 \ell^4 r^{10} + 115166607835136 \ell^2 r^{12} + 173095839072256 r^{14} \big)\nonumber\\
 g_{4;4}={}&\frac{1}{6138850443264000}\big(712893 \ell^{18} + 136875456 \ell^{16} r^2 + 12045040128 \ell^{14} r^4 + 642402140160 \ell^{12} r^6\nonumber\\
 & + 23033726566400 \ell^{10} r^8 + 582650293125120 \ell^8 r^{10} + 10646216119418880 \ell^6 r^{12} \nonumber\\
 &+ 141874001126359040 \ell^4 r^{14} + 885036208147660800 \ell^2 r^{16} + 1854106457918668800 r^{18} \big)\nonumber\\
 \Omega_{4;4}={}&\frac{1}{2302068916224000}\big(2823349 \ell^{20} + 496909424 \ell^{18} r^2 + 39752753920 \ell^{16} r^4 + 1908132188160 \ell^{14} r^6\nonumber\\
 & + 60702439243776 \ell^{12} r^8 + 1331841818689536 \ell^{10} r^{10} + 20358704670965760 \ell^8 r^{12} \nonumber\\
 &+ 214664792789483520 \ell^6 r^{14} + 1320189581475184640 \ell^4 r^{16} + 4227781500545794048 \ell^2 r^{18} \nonumber\\
 &+ 6149500364430245888 r^{20} \big)\nonumber\\
\Pi_{4;5}={}&\frac{1}{2707728082379168808960000}\big(359856139756087 \ell^{22} + 68759317607951424 \ell^{20} r^2\nonumber\\
& + 6018441511070587392 \ell^{18} r^4 + 319055812097689436160 \ell^{16} r^6 + 11522040451970846883840 \ell^{14} r^8\nonumber\\
& + 300448881088844399640576 \ell^{12} r^{10} + 5766531014229801704620032 \ell^{10} r^{12} \nonumber\\
&+ 80501048159287373155270656 \ell^8 r^{14} + 813902149229647850593320960 \ell^6 r^{16} \nonumber\\
&+ 5046507556181843886273986560 \ell^4 r^{18} + 16003427123832136843395072000 \ell^2 r^{20} \nonumber\\
&+ 19843710569268281812671528960 r^{22}\big)\nonumber
\end{align}

\bibliographystyle{plain}

\end{document}